\documentclass[aps,prb,twocolumn,floats,showpacs,superscriptaddress,longbibliography]{revtex4-1}

\usepackage{bm}
\usepackage{graphicx}
\usepackage{epstopdf}
\usepackage{amsmath}
\usepackage{times}

\def \FUW{Institute of Experimental Physics, Faculty of Physics,
University of Warsaw, Pasteura 5, 00-093 Warsaw, Poland}

\def \IFPAN{Institute of Physics, Polish Academy of Sciences, al. Lotnik\'{o}w 32/64, 02-688 Warsaw, Poland}

\def \BREMA{Institute of Solid State Physics, Semiconductor Epitaxy, University of Bremen, PO Box 330 440, D-28334 Bremen, Germany}

\def \GRENOBLE{Laboratoire National des Champs Magn\'etiques Intenses CNRS-UGA-UPS-INSA-EMFL, 30942 Grenoble, France}

\begin{document}

\title{Comparison of magneto-optical properties of various excitonic complexes in CdTe and CdSe self-assembled quantum dots}
\author{J.~\surname{Kobak}}\email{Jakub.Kobak@fuw.edu.pl}\affiliation{\FUW}
\author{T.~\surname{Smole\'nski}}\affiliation{\FUW} 
\author{M.~\surname{Goryca}}\affiliation{\FUW} 
\author{J.-G.~\surname{Rousset}}\affiliation{\FUW} 
\author{W.~\surname{Pacuski}}\affiliation{\FUW}
\author{A.~\surname{Bogucki}}\affiliation{\FUW} 
\author{K.~\surname{Oreszczuk}}\affiliation{\FUW} 
\author{P.~\surname{Kossacki}}\affiliation{\FUW}
\author{M.~\surname{Nawrocki}}\affiliation{\FUW}
\author{A.~\surname{Golnik}}\affiliation{\FUW} 
\author{J.~\surname{P\l{}achta}}\affiliation{\IFPAN}
\author{P.~\surname{Wojnar}}\affiliation{\IFPAN}
\author{C.~\surname{Kruse}}\affiliation{\BREMA}
\author{D.~\surname{Hommel}}\affiliation{\BREMA}
\author{M.~\surname{Potemski}}\affiliation{\GRENOBLE}
\author{T.~\surname{Kazimierczuk}}\affiliation{\FUW} 

\begin{abstract} 
We present a comparative study of two self-assembled quantum dot (QD) systems based on II-VI compounds: CdTe/ZnTe and CdSe/ZnSe. Using magneto-optical techniques we investigated a large population of individual QDs. The systematic photoluminescence studies of emission lines related to the recombination of neutral exciton X, biexciton XX, and singly charged excitons (X$^+$, X$^-$) allowed us to determine average parameters describing CdTe QDs (CdSe QDs): X--XX transition energy difference $12$~meV ($24$~meV); fine-structure splitting \mbox{$\delta_{1}=0.14$~meV} \mbox{($\delta_{1}=0.47$~meV)}; $g$-factor $g=2.12$ ($g=1.71$); diamagnetic shift \mbox{$\gamma=2.5~\mu$eV$/$T$^{2}$} \mbox{($\gamma=1.3~\mu$eV$/$T$^{2}$)}. We find also statistically significant correlations between various parameters describing internal structure of excitonic complexes.

\end{abstract}



\maketitle

\section{Introduction}

Epitaxial quantum dots (QDs) are renowned for their diversity --- in a single sample one can find QDs with different values of emission energy, anisotropy-induced exchange splitting,  effective Land\'e factor,
diamagnetic shift, and other parameters. It can be an advantage, if a single QD with particular properties (e.g., zero anisotropy splitting \cite{young-prb-2005, stevenson-nature-2006}) is required. On the other hand,
such a diversity is an obstacle on the way to determine the \emph{typical} behavior of QDs in a given material system.

Precise determination of a typical QD parameters requires averaging over many individual dots. In case of some characteristics, such as emission energy or $g$-factor, it is possible to simply measure the response of the whole QD ensemble,\cite{Zhang-JVSTB-2010, Syperek-PRB-2011, Reshina-PRB-2012, Man-SR-2015} e.g., in photoluminescence (PL) or time-resolved Faraday rotation experiments, respectively. However, more detailed characteristics such as anisotropic fine-structure splitting (FSS) can be studied directly only on a single-dot level and thus a significant number of individual dots need to be analyzed in order to draw robust conclusions about the average value.

In this work we present results of systematic comparison between the two popular II-VI self-assembled QD systems: CdTe/ZnTe and CdSe/ZnSe. We particularly focus on differences between excitons of various charge states: their binding energy, $g$-factor and diamagnetic shift. These quantities have been already measured for single quantum dots \cite{walck-prb-1998,Kulakovskii-PRL-1999,Hundt-PSSB-2001,bayer-prb-2002,besombes-prb-2002,finley-prb-2002,schulhauser-pss-2003,Akimov-PRB-2005,Leger-PRB-2007,oberli-prb-2009,Hewaparakrama-NT-2008}, but have not been analyzed in terms of variation across the QD population.

\section{Samples and experimental setup} 

We studied 3 structures with CdSe QDs in ZnSe barriers and 4 structures with CdTe QDs in ZnTe barriers. Samples were fabricated in three different laboratories (affiliations 1, 2, and 3). The growth of the structures was performed by molecular beam epitaxy (MBE) on GaAs substrates. In most cases the reorganization of the QDs was induced by a well-established amorphous Te or Se desorption method \cite{tinjod-apl-2003,wojnar-jcg-2011,kobak-JCG-2013} for which the growth temperature was varied just after deposition of the QD formation layer. For the two selenide samples the cap layer was deposited directly on the QDs layer without changing the substrate temperature. In order to reduce QDs density in selenide samples we applied additional low level delta-doping with transition metal ions \cite{kobak-nature-2014,smolenski-prb-2015orientation,Smolenski-NC-2016}. However, in this work we include only results obtained for individual QDs, which do not contain magnetic ions inside.

The studied samples were placed inside a magneto-optical bath cryostat with magnetic field of up to 10~T. The measurements were performed at the temperature of about 1.5~K using a reflective type microscope, which focuses the laser beam to a 0.5~$\mu$m diameter spot. This allowed us to study optical properties of well-resolved emission lines of single QDs in high magnetic fields with a polarization resolution. Complementary magneto-optical studies were performed in Grenoble High Magnetic Field Laboratory, where a helium-bath cryostat (4.2~K) with a sample was placed inside a 20~MW resistive magnet producing magnetic field of up to 28~T.

\section{Results}

In order to present statistically significant data we have investigated over 160 individual QDs. For each analyzed QD we studied emission lines originating from the recombination of the neutral exciton X, the biexciton XX, and the charged excitons (X$^+$, X$^-$). Identification of such lines has been already discussed in detail elsewhere.\cite{Kulakovskii-PRL-1999,besombes-prb-2002,Suffczynski-PRB-2006,kazimierczuk-prb-2011-exchange,Kazimierczuk-PRB-2013} Typically it includes the analysis of linear polarization of emission and the dependence of the PL intensity on the excitation power. Based on the PL spectra measured in magnetic field of 0-10~T we extracted parameters describing each of the studied excitonic transitions: their relative emission energies, anisotropy-induced exchange splittings, effective Land\'e $g$-factors, and diamagnetic shift coefficients $\gamma$. 

\begin{figure}
\includegraphics[width=1\linewidth]{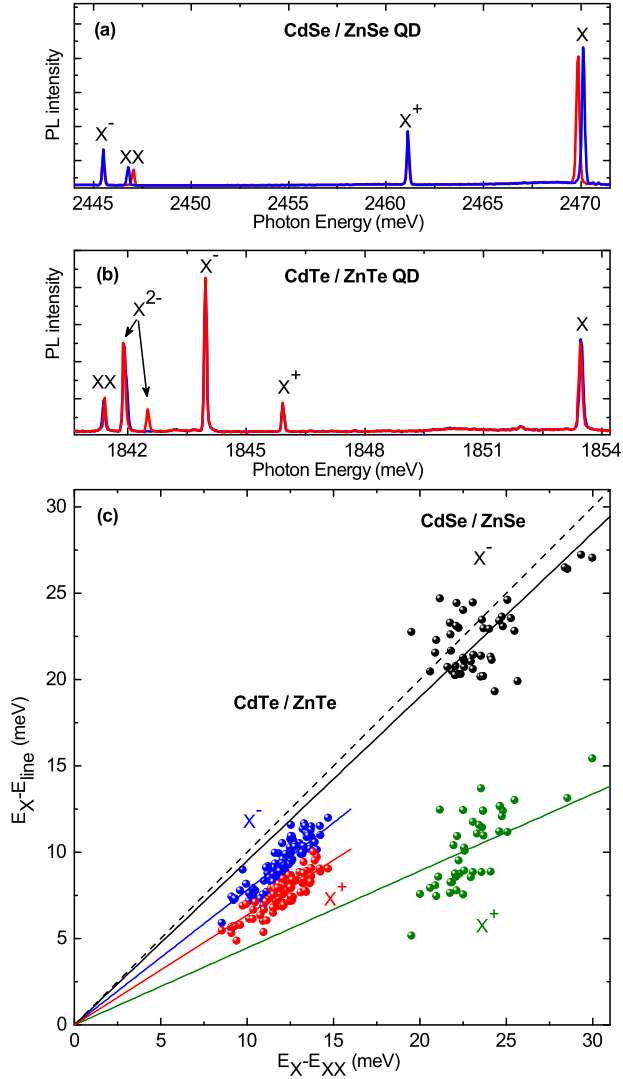}
\caption{(Color online) Relative emission energy of various excitonic complexes in QDs. (a,b) Photoluminescence spectra of CdTe/ZnTe and CdSe/ZnSe QDs, respectively. The QDs show typical anisotropy properties: the emission lines of the neutral exciton (X) and biexciton (XX) exhibit opposite linear polarizations, while the lines related to the trions (X$^+$,~X$^-$) are not linearly polarized. (c) Energy difference between the neutral exciton line and the charged exciton lines plotted versus the energy difference between the neutral exciton line and the biexciton line for CdTe/ZnTe (red and blue symbols) and CdSe/ZnSe (black and green symbols) QDs. Solid lines mark the linear fits ($y = ax$) with proportionality constants equal to $a_{\mathrm{X^{+}}}^{\mathrm{CdTe}}=0.64$, $a_{\mathrm{X^{-}}}^{\mathrm{CdTe}}=0.78$, $a_{\mathrm{X^{+}}}^{\mathrm{CdSe}}=0.45$, $a_{\mathrm{X^{-}}}^{\mathrm{CdSe}}=0.95$ (compare with Refs. \onlinecite{kazimierczuk-prb-2011-exchange,kobak-APPA-2011}). Dashed line corresponding to $y=x$ is drawn for the reference.}
\label{CdTeAndCdSeQDs_PolozeniaLinii}
\end{figure}
 
\subsection {Relative emission energy of  X, XX, X$^{+}$, and X$^{-}$}

The energy of the emission lines related to particular QD strongly depends, e.g., on the growth procedure and the QD size \cite{kobak-JCG-2013}, but in this work we focus rather on the structure of a PL spectrum of a single QD. Figs~\ref{CdTeAndCdSeQDs_PolozeniaLinii}(a) and \ref{CdTeAndCdSeQDs_PolozeniaLinii}(b) present PL spectra of a single CdSe/ZnSe and CdTe/ZnTe QD. Due to the Coulomb interaction, emission lines of different excitonic complexes are shifted from the basic neutral exciton X. Relative energies of different lines may vary depending on the QD size, shape and composition.\cite{Zielinski-PRB-2015} However, in accordance with Ref. \onlinecite{kazimierczuk-prb-2011-exchange,kobak-APPA-2011} we find  that for telluride QDs the distances between the emission lines vary almost proportionally to each other (Fig.~\ref{CdTeAndCdSeQDs_PolozeniaLinii}(c)). Consequently, the emission pattern stays roughly the same, except for some variation of the horizontal scale. Such an effect significantly simplifies identification of the PL lines originating from a single CdTe/ZnTe QD in the experiment. As shown in Fig.~\ref{CdTeAndCdSeQDs_PolozeniaLinii}(c), the situation in the selenide system is different. Distances between the X$^+$, X$^-$ and X lines in case of the selenide QDs cannot be considered proportional to the X--XX distance. Although there exists some positive correlation between the \mbox{X--X$^-$}, X--X$^+$ distances and the X--XX distance (Pearson's correlation coefficients $r_{\mathrm{X}^{-}}=0.58$ and $r_{\mathrm{X}^{+}}=0.66$, compared to $r_{\mathrm{X}^{-}}=0.87$ and $r_{\mathrm{X}^{+}}=0.87$ found for the tellurides), their relationship does not correspond to a simple proportionality (see Fig.~\ref{CdTeAndCdSeQDs_PolozeniaLinii}(c)). We note that while all studied CdTe QDs exhibited the same sequence of excitonic transitions in the PL spectrum, for the selenide QDs the negatively charged exciton line can be situated either on higher or on lower energetic side of the neutral biexciton line. 

Another important difference between the two material systems is the average value of the relative X--XX transition energy distance. In the case of telluride QDs such an average energy distance is equal to about $12.0$~meV, while for the selenide QDs the obtained average value is about twice as high and yields $23.9$~meV.

\subsection {Anisotropic exchange splitting of X and XX}

\begin{figure}
\includegraphics[width=1\linewidth]{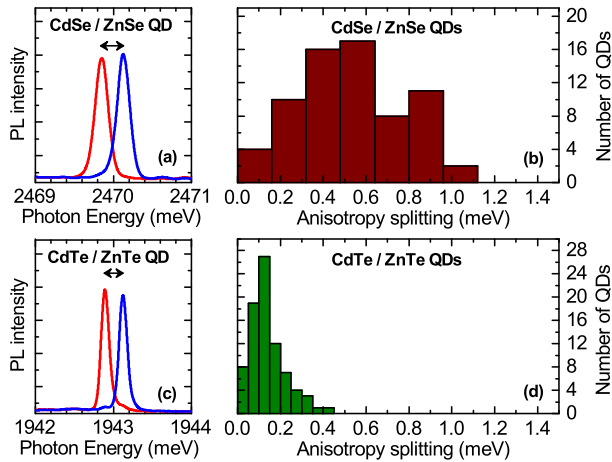}
\caption{(Color online) Anisotropy splitting of the neutral exciton. (a, c) Example PL spectra of the bright neutral exciton (X) in CdSe/ZnSe and CdTe/ZnTe QD, respectively. The spectra were detected in two orthogonal linear polarizations, the orientations of which correspond to the principal axes of the QD anisotropy. (b, d) Distribution of the X anisotropy splitting in various CdSe/ZnSe and CdTe/ZnTe QD, respectively.}
\label{CdTeAndCdSeQDs_anizotropia}
\end{figure}

By means of the PL measurements performed with a polarization resolution of detection we analyzed the fine-structure of the X and XX states for CdTe and CdSe QDs (Fig.~\ref{CdTeAndCdSeQDs_anizotropia}). In the case of both excitonic complexes the zero-field emission contains two lines split by the energy related to the anisotropic part of the exchange interaction between the electron and the heavy hole \cite{bayer-prb-2002}. Such emission lines can be seen separately in two perpendicular linear polarizations of the detection. As expected, for each QD we observed the same value of anisotropy splitting of X and XX, but opposite ordering of the fine-structure-split components for both complexes. This is due to the fact that X is the final state of the XX transition. In the experiment, for each studied dot we collected the PL spectra as a function of linear polarization angle of detection. Such a measurement provides information not only about the value of the fine-structure splitting ($\delta_{1}$) but also about the in-plane anisotropy axis of each QD. Coherently with the previous reports on II-VI QDs \cite{kazimierczuk-prb-2011-exchange,kudelski-icps} we do not observe any correlation between the in-plane anisotropy parameters (splitting and direction) and both the transition energy and the biexciton relative energy. For the telluride QDs the average value of the fine-structure splitting $\delta_{1}$ is about $0.14$ meV, whereas for the selenide QDs we found a few times higher average value of $\delta_{1} = 0.47$ meV.

\subsection {Zeeman effect and diamagnetic shift}

The energy spectrum of the QD can be manipulated by external magnetic field. Two main effects occurring upon application of the magnetic field are a linear splitting of the states depending on the spin projection (Zeeman effect)\cite{bayer-prb-2002,besombes-prb-2002,schulhauser-pss-2003} and a quadratic energy shift due to a finite spatial extension of the exciton wave function (diamagnetic shift)\cite{walck-prb-1998,schulhauser-pss-2003}. The strength of these effects is parametrized with the excitonic $g$-factor and the diamagnetic field coefficient $\gamma$. 

In our experiments we studied these parameters in the Faraday geometry with the magnetic field applied along the growth axis of the QDs. As expected, the transitions of all considered excitonic complexes exhibit a common field-induced splitting pattern comprising of two lines separated by the energy of $\sqrt{\Delta^2+(g\mu_BB)^2}$, where $g$ is the $g$-factor, while $\Delta$ corresponds to the zero-field splitting (equal to $\delta_1$ for the neutral complexes and 0 in case of the trions). In order to take into account the change of the mean emission energy (originating from the diamagnetic shift), we fitted both Zeeman branches of a given transitions with a quadratic formulas of the form $E(B) = E_{0} \pm \frac{1}{2}\sqrt{\Delta^2+(g\mu_{B}B)^2} + \gamma B^{2}$, where $+$ ($-$) sign corresponds to the higher (lower) energy line. 

\begin{figure*}
\includegraphics[width=1\linewidth]{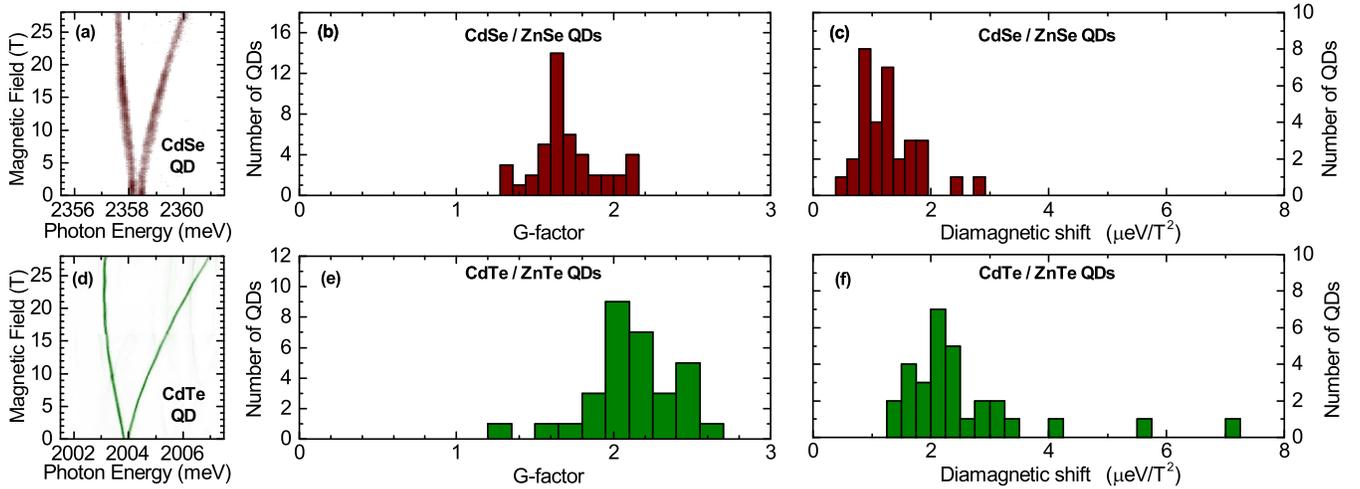}
\caption{(Color online) Magneto-spectroscopy of the neutral exciton state (X) together with the distribution of excitonic $g$-factors and diamagnetic shift coefficients among the studied dots. (a, d) The evolution of X emission line in external magnetic field applied in the Faraday geometry. (b, e) Histograms of the Land\'e $g$-factor values for various CdSe/ZnSe and CdTe/ZnTe QDs. (c, f)
Histograms of diamagnetic shift coefficients for various CdSe/ZnSe and CdTe/ZnTe QDs. }
\label{CdTeAndCdSeQDs_GczynnikiIDiamShift}
\end{figure*}

The results of the fitting are presented in Fig. \ref{CdTeAndCdSeQDs_GczynnikiIDiamShift}. The average $g$-factor of the CdTe QDs ($g=2.12$) was found to be slightly larger than the average $g$-factor of the CdSe QDs ($g=1.71$). More significant differences we obtained by studying the diamagnetic shift, which was found to be approximately two times higher for the telluride QDs ($2.5~\mu$eV$/$T$^{2}$) compared to the selenide ones ($1.3~\mu$eV$/$T$^{2}$), as shown in Figs~\ref{CdTeAndCdSeQDs_GczynnikiIDiamShift}(c,f). These values can be expressed using a more comprehensive quantity of the spatial extension of the excitonic wave function according to a relation
\begin{equation}
\gamma = \frac{\mathrm{e}^2}{8 m} \langle r^2 \rangle,
\end{equation}
where $m$ is the in-plane reduced mass of the exciton. One should note that this expression is strictly valid only for the systems with translational symmetry in the plane perpendicular to the magnetic field (i.e., in bulk or quantum wells) 
\cite{walck-prb-1998}. By applying this formula also to the case of QDs we obtained the values of the spatial extension of the excitonic wave function in normal configuration for CdSe and CdTe QDs as $\sqrt{\langle r^2\rangle}=2.4$~nm and $3.1$~nm, respectively. Such values stay in a good agreement with the material trends. More specifically, dielectric constant is smaller for the selenides\cite{Strzalkowski-APL-1976}, which leads to a stronger electron-hole interaction. As a consequence, the CdSe QDs exhibit stronger exciton binding and smaller radius of the excitonic wave function. The obtained values can be also compared with the bulk exciton radii, which yield 5.6~nm for CdSe and 7.5~nm for CdTe\cite{Landolt_Bohr_Radius}. In each system the confinement in the QD potential reduces the extension of the exciton wavefunction, yet only up to about 60\%.

\begin{figure}
\includegraphics[width=1\linewidth]{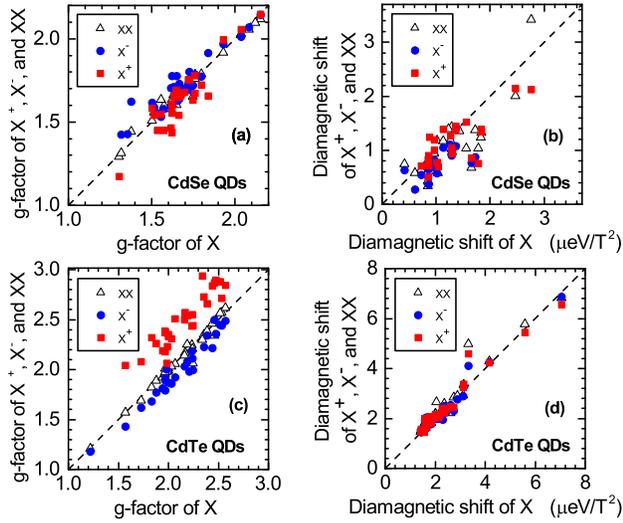}
\caption{(Color online) Statistic parameters describing the magneto-optical characteristics of CdSe/ZnSe (a, b) and CdTe/ZnTe (c, d) QDs in the Faraday configuration: (a, c) Correlation of the Land\'e $g$-factors for different excitonic complexes. (b, d) Correlation of diamagnetic shift coefficients for different excitonic complexes.}
\label{CdTeAndCdSeQDs_GczynnikiEkscytonow}
\end{figure}

The most surprising findings are obtained from the correlations between the effective Land\'e $g$-factors determined for different excitonic complexes. Since each of the studied lines is related to the recombination of an s-shell hole with an \mbox{s-shell} electron, all these transitions for a given QD are expected to exhibit the same excitonic $g$-factor. Nevertheless, Y. L\'eger \emph{et al.}\cite{Leger-PRB-2007} reported about a CdTe QD with different values of $g$-factors for various excitonic complexes. Our measurements corroborate this claim and demonstrate the existence of a systematic deviation between the $g$-factors of X, X$^+$, and X$^-$ transitions in the whole population of the QDs. Such a systematic difference was found in both systems, but with opposite sign, as seen in Fig.~\ref{CdTeAndCdSeQDs_GczynnikiEkscytonow}. 

In the case of CdTe QDs we observed with perfect regularity that the $g$-factor of the neutral exciton is greater than the \mbox{$g$-factor} of the X$^-$ and smaller than the $g$-factor of the X$^+$. The average difference between the $g$-factor values corresponding to X$^+$ and X was equal to $0.34$, while the difference between the $g$-factor of X$^-$ and X was about three times smaller ($0.11$). For the selenide QDs we observed opposite sign of this effect: greater values of the $g$-factors for X$^-$ states and smaller for X$^+$ stats in comparison to X Land\'e factor. However, due to significantly lower mean values of the $g$-factor differences and fluctuations of the g-factor distribution we found a few exceptions from such an ordering. With respect to the X $g$-factor, Land\'e factor of X$^-$ state was on average larger by $0.03$, while the $g$-factor of X$^+$ state was smaller by about $0.04$. We interpret the observed differences between the $g$-factor values for various excitonic complexes as resulting from significant modification of the carrier wave function imposed by the presence of the other carriers in the dot.\cite{Leger-PRB-2007}

We note that for both QDs systems the Land\'e factors of X and XX were equal for all dots within our experimental uncertainty. Such a relation is expected, since the singlet nature of the biexciton state implies that the Zeeman effect of the XX transition originates solely from the final state, i.e., the neutral exciton state. Average differences of the $g$-factors of X and XX states were equal $0.006$ and $0.007$ for telluride and selenides QDs, respectively, which is indeed much smaller than differences between the charged excitons $g$-factors discussed earlier.

In analogy to the study of the Land\'e factors, we also investigated correlations between diamagnetic shift constants for various excitonic complexes. For both QDs systems we do not observe any systematic differences, except that for selenide QDs the diamagnetic shift of the X tended to be slightly larger than for the other complexes. However, in the case of CdTe QDs we obtained similar values of the diamagnetic shifts coefficients for various excitons in a given dot.

\section {Summary}

By means of magneto-optical techniques we have studied and compared the two systems of self-organized QDs: CdTe/ZnTe and CdSe/ZnSe QDs. We investigated over 160 randomly selected individual QDs. To reduce the influence of the effects related to the growth technique and specific sample, we examined 7 structures fabricated in 3 laboratories. Based on such statistical approach we determined the key parameters describing magneto-optical properties of the excitonic complexes in CdTe/ZnTe and CdSe/ZnSe QDs. The average values of the characteristic parameters describing the studied QDs are summarized in Table~\ref{CdTeAndCdSeQDs_parametersTable}.

\begin{table}[b]
\begin{tabular}{|l|c|c|} 
\hline
& CdSe/ZnSe QDs & CdTe/ZnTe QDs \\
\hline
$E_\mathrm{X}-E_\mathrm{XX}$ (meV) & $23.9 \pm 2.6$ & $12.0 \pm 1.2$ \\
$\delta_{1}$ (meV) & $0.47 \pm 0.21$ & $0.14 \pm 0.08$ \\
\hline
$g_\mathrm{X}$ (number) & $1.71 \pm 0.21$ & $2.12 \pm 0.30$ \\
$g_\mathrm{X^{+}}-g_\mathrm{X}$ (number) & $-0.04 \pm 0.07$ & $0.34 \pm 0.11$ \\
$g_\mathrm{X^{-}}-g_\mathrm{X}$ (number) & $0.03 \pm 0.08$ & $-0.11 \pm 0.07$ \\
\hline
$\gamma_\mathrm{X}$ ($\mu$eV$/$T$^{2}$) & $1.3 \pm 0.5$ & $2.5 \pm 1.2$ \\
\hline
\end{tabular}
\caption{Average values of the parameters describing CdTe and CdSe QDs. The symbols are introduced and explained in the text. Uncertainties were calculated as a standard deviation of the determined parameters.}
\label{CdTeAndCdSeQDs_parametersTable}
\end{table}

Typical spectra of individual CdTe QD contain several emission lines that form a characteristic pattern. For most of the studied CdTe dots the energy distances between the emission lines related to various charged states vary proportionally to the energy distance of X and XX. In the case of selenide QDs we observed much weaker correlation of the relative energy positions of the emission lines. Furthermore, the sequence of the emission lines is not conserved for all dots, i.e., the energy of X$^-$ emission line can be either higher and lower than the energy of XX line. Analysis of the magnetic field dependence of the emission spectra revealed an unexpected effect of the systematic difference between the $g$-factors of various excitonic complexes, especially for the CdTe system.

The comparison between the parameters determined for CdTe and CdSe QDs is consistent with the general material trends. In particular, the smaller dielectric constant of selenides\cite{Strzalkowski-APL-1976} leads to a stronger Coulomb interaction between the carriers, which is reflected by the higher bulk exciton binding energy.\cite{Voigt-PSSB-1979,Nawrocki-PSSB-1980,Wagner-PhysB-1993,Aliev-PSS-1994, Worz-PSSB-1997} Similarly, the selenide QDs exhibit larger energetic distance between the exciton and the biexciton emission lines, larger anisotropic part of the electron-hole exchange interaction and tighter binding of the carriers in the excitonic complexes evidenced by a smaller value of the diamagnetic shift coefficient. 

\begin{acknowledgments} This work was partially supported by the Polish
National Science Centre under decisions DEC-2012/05/N/ST3/03209, DEC-2013/09/B/ST3/02603, DEC-2011/02/A/ST3/00131,  and by Polish Ministry of Science and Higher Education in years 2012$-$2017 as research grants ``Diamentowy Grant'' and Iuventus Plus (N$^{\circ}$ IP2014 034573). One of us (T.S.) was supported by the Foundation for Polish Science through the START programme. The research leading to these results has received funding from the European Union Seventh Framework Programme (FP7/2007-2013) under grant agreement N$^{\circ}$ 316244. Project was carried out with the use of CePT, CeZaMat, and NLTK infrastructures financed by the European Union - the European Regional Development Fund within the Operational Programme "Innovative economy".
\end{acknowledgments}

\end{document}